\documentclass[twocolumn]{aastex631}
\PassOptionsToPackage{hyphens}{url}
\usepackage[hyphens]{url}
\usepackage{breakurl}

\newcommand{\eiso}{$E_{\rm iso}$}
\newcommand{\liso}{$L_{\rm iso}$}
\newcommand{\epo}{$E_{\rm p,o}$}
\newcommand{\epi}{$E_{\rm p,i}$}

\shorttitle{peak flux in LGRB fundamental diagram}
\shortauthors{Lan et al.}
\graphicspath{{./}{figures/}}

\begin{document}

\title{The peak flux of long gamma-ray bursts  in the E$_{\rm p,i}$--L$_{\rm iso}$ diagram and in the classification}

\correspondingauthor{Guangxuan Lan}
\email{gxlan@pku.edu.cn}

\author{Guangxuan Lan}
\affiliation{Department of Astronomy, School of Physics, Peking University, Beijing, China}
\affiliation{IRAP, Universit\'{e} de Toulouse, CNRS, CNES, UPS, Toulouse, France}
\author[0000-0001-7346-5114]{Jean-Luc Atteia}
\affiliation{IRAP, Universit\'{e} de Toulouse, CNRS, CNES, UPS, Toulouse, France}



\begin{abstract}
The \epi--\liso\ correlation of long gamma-ray bursts (LGRBs) is regarded as a fundamental correlation for standardizing LGRBs to probe cosmology and constrain LGRB physics. However, this correlation may be affected by potential selection effects which are likely overlooked in the current small LGRB redshift sample. In this work, we simulate a large LGRB sample that reflects the observed situation, aiming to study the impact of peak flux $P$ on the observed LGRB \epi--\liso\ correlation. We find that the overall (\epi, \liso) distribution, which will directly affect the best-fit result of the correlation, is significantly dependent on the value of $P$. This indicates that the impact of peak flux selection should be carefully considered in the studies and applications of the \epi--\liso\ correlation. Notably, we show that our simulated data can reproduce the observed $P$ distribution only if some dependence of (\epi, \liso) is included in the simulation. This is an indication that the (\epi, \liso) connection is a crucial property of LGRBs. We also find that GRBs with high peak flux in the low-\epi\ \& \liso\ region are not the straightforward extrapolation of the GRB population in the higher-\epi\ \& \liso\ region. Selecting four bursts with $ L_{\rm iso}\le10^{50}$ erg s$^{-1}$, $E_{\rm p, i}\le10^{2.5}$ keV, and $P\ge10^{0.5}$ ph cm$^{-2}$ s$^{-1}$, we find two bursts, GRB~060614 and GRB~191019A, which may not be associated with the theoretical massive-star origin of LGRBs. This suggests that combining $P$ with the position in the \epi--\liso\ diagram may be used to indicate alternative LGRB origins.

\end{abstract}

\keywords{Gamma-ray bursts (629)}


\section{Introduction}
\label{sec:intro}

Long gamma-ray bursts (LGRBs, GRBs with duration greater than 2 s) are extremely luminous transient sources shining in the (sub) gamma-ray band in the universe. There is a consensus to attribute them to transient ultra-relativistic jets emitted in our direction by newly born compact objects formed during the collapse of the core of the massive star \citep{2006ARA&A..44..507W}. Thanks to their luminosity, LGRBs can be detected out to very large distances, therefore constituting a potential tool to probe the early universe. After the finding of some empirical correlations like the \epi -- \liso\ correlation \citep{2004ApJ...609..935Y} and the \epi --\eiso\ correlation \citep{2002A&A...390...81A}, which show the intrinsic LGRB nature between the rest-frame LGRB spectral peak energy and LGRB isotropic-equivalent luminosity (energy), LGRBs have been widely employed to expand the Hubble diagram and constrain cosmological parameters \citep[see][and the references therein]{2015NewAR..67....1W,2018PASP..130e1001D,2019MNRAS.486L..46A}. These correlations are also used to study the LGRB physics \citep{2019NatCo..10.1504I}, LGRB formation rate \citep{2016A&A...587A..40P,2019MNRAS.488.4607L,2021MNRAS.508...52L,2022ApJ...938..129L,2021A&A...649A.166P}, and to estimate the pseudo redshifts of LGRBs \citep{2003A&A...407L...1A,2004ApJ...609..935Y,2013ApJ...772L...8T,2018Ap&SS.363..223Z}.

However, the existence of some outliers \citep{2005ApJ...627..319B,2005MNRAS.360L..73N,2013A&A...557A.100H} and the impact of instrumental effects \citep{2007ApJ...671..656B,2010ApJ...711..495B,2011MNRAS.411.1843S,2012ApJ...747...39C,2012ApJ...747..146K,2013ApJ...766..111S} weaken the applicability of these correlations. For example, by giving a redshift in the range [0.34-4.35] to \textit{Fermi} bursts without measured redshift, \cite{2013A&A...557A.100H} found that many bursts will be outside the 2$\sigma$ region of the \epi--\eiso\  correlation obtained from bursts with redshift measured in the same range, no matter the redshift considered for them. Simultaneously, these outliers have smaller average fluence and peak flux than the bursts within the 2$\sigma$ \epi--\eiso correlation. This inspires us to investigate the potential selection effects in the \epi--\liso\ and \epi--\eiso\ correlations because the observed GRB sample is unavoidably affected by the instrumental limits \citep{2003ApJ...588..945B} and may also be constrained by sample selection criteria \citep{2012ApJ...749...68S,2012MNRAS.421.1256N,2016ApJ...817....7P,2016A&A...587A..40P,2021A&A...649A.166P} in fluence or peak flux. Confirming whether such selection effects will bias the best-fit results of the spectral--energy (luminosity) correlations is crucial for their utilization.

The impact of instrumental sensitivity on the observed distributions of \epi, \liso, and \eiso, and potentially on the validity of the spectral correlations was studied by various authors \citep{2007ApJ...671..656B,2008MNRAS.391..639N,2010ApJ...711..495B,2013ApJ...766..111S,2018PASP..130e1001D}. Although they can explain the limit and scatter of the observed spectral correlations, how the selection effects affect the best-fit results of the correlations has been rarely addressed. This is the aim of this work. We consider here the \epi--\liso\ correlation of LGRBs as an example to discuss the impact of peak flux selection on the (\epi , \liso\ ) distribution and on the best-fit result of the correlation.

This paper is organized as follows: the methodology is presented in Section \ref{sec:sim}. The impact of selection effects on the \epi--\liso\ distribution is discussed in Section \ref{sec:result}, along with other effects like the boundary of the joint distribution in the low-\epi\ \& high-\liso\ region and the redshift evolution of the \epi--\liso\ correlation. Finally, we summarize our work and findings in Section \ref{sec:sum}. Throughout this paper, a flat $\Lambda$CDM cosmological model with
$H_{0}=70$ km s$^{-1}$ Mpc$^{-1}$, $\Omega_{\rm m}=0.3$, and $\Omega_{\Lambda}=0.7$ is adopted  like previous works \citep{2016A&A...587A..40P,2021MNRAS.508...52L}. The best-fit lines are all obtained with an orthogonal regression method\footnote{\url{https://www.mathworks.com/matlabcentral/fileexchange/16800-orthogonal-linear-regression}} that can take into account the uncertainties on both x- and y- axes data.

\section{Methodology and Sample Construction}
\label{sec:sim}

\begin{figure}
\center
\includegraphics[angle=0,scale=0.6]{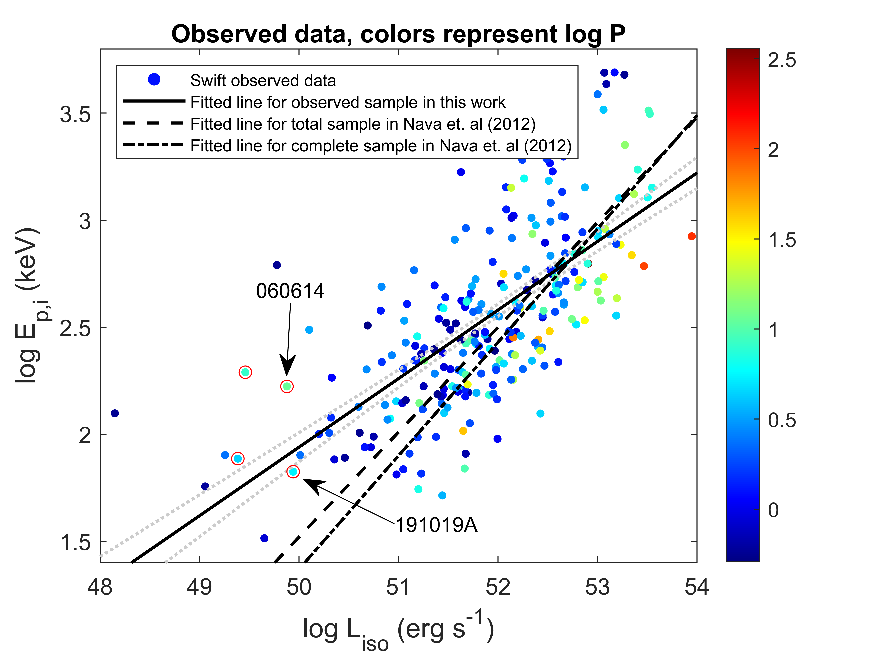}
\caption{The \epi\ vs. \liso\ distribution for 259 \textit{Swift} observed GRBs with a measured redshift. The colors represent the different values of $\log P$. The three fitted lines are best-fit results for the \epi--\liso\ correlation based on the sample in this work (solid), the total sample (dashed) and the complete sample (dash-dotted) in \cite{2012MNRAS.421.1256N}, respectively. The dotted lines are 1$\sigma$ confidence interval of the solid line. The red circles are 4 LGRBs with $ L_{\rm iso}\le10^{50}$ erg s$^{-1}$, $E_{\rm p, i}\le10^{2.5}$ keV, and $P\ge10^{0.5}$ ph cm$^{-2}$ s$^{-1}$.
\label{fig:obsEpiL}}
\end{figure}

Following other works in this domain, we base our study on a simulated GRB population, whose construction is described below. This choice is motivated by the relatively small number of GRBs with a redshift and E$_{\rm peak}$ measurement. The difficulty to work with an observed sample is illustrated in Fig. \ref{fig:obsEpiL}, which shows the observed \epi--\liso\ diagram with colors that reflect the value of $\log P$. Even if the impact of peak flux is not significant here due to the small sample size, the best-fit \epi--\liso\ relation differs from the calculation of \cite{2012MNRAS.421.1256N} for their total sample (all bursts with $z$) and their complete sample ($P>2.6$ ph cm$^{-2}$ s$^{-1}$). Since the fitting results of the \epi--\liso\ correlation are dependent on the selected sample, we are led to consider that selection effects impact the correlation but the number of LGRBs with a measured redshift is too sparse to clearly reveal this impact. Therefore, we base our study on a large sample of 10,000 simulated LGRBs that closely reproduce the main properties of the observed long GRB population. Such a large population allows to better evidence the impact of selection effects.

\subsection{Sample Construction}
\label{ssec:sample}

To study the \epi--\liso\ correlation, a large number of GRBs with measured redshifts is necessary. So far, only the \textit{Swift} satellite has a significant number of GRB redshifts in reserve. This determines the observational sample used for this work. Even so, to clearly reveal the selection effects in the \epi--\liso\ correlation, particularly how peak flux $P$ affects the (\epi, \liso) distribution that we are interested in here, a population larger than the current LGRB number is still required. To achieve this aim, we generate a large synthetic population from the GRB world model of \cite{2021MNRAS.508...52L}, which includes the quantities $z$, $P$, $\alpha$, \epo, and $L_{\rm iso}$ that can represent the observed sample. The simulated results of these quantities will be compared with the observed results ensuring that the mock sample reflects the real situation. Since the simulation needs the LGRBs' spectra to be coherent with the observations, we employ the parameters from the cut-off power-law (CPL) model for the observed LGRBs. This model has been used by the \textit{Swift} team to have the time-integrated spectral fit for most of the spectra of LGRBs detected with BAT. The best-fit peak flux $P$ (ph cm$^{-2}$ s$^{-1}$), the index of the CPL spectral model $\alpha$, and the peak energy $E_{\rm p,o}$ (keV) computed from the spectrum in the energy range 15--150 keV can be downloaded directly from the \textit{Swift}/BAT Gamma-Ray Burst Catalog\footnote{\url{https://swift.gsfc.nasa.gov/results/batgrbcat/index_tables.html}}. With these parameters, we can finally obtain the rest-frame LGRB peak energy $E_{\rm p,i}=E_{\rm p,o}(1+z)$ and the isotropic-equivalent luminosity $L_{\rm iso}=4\pi D_L^2(z) P \frac{\int^{10^4/(1+z)\;{\rm keV}}_{1/(1+z)\;{\rm keV}} EN(E)dE}{\int^{150\;{\rm keV}}_{15\;{\rm keV}} N(E)dE}$, where $D_L(z)$ is the luminosity distance at redshift $z$ and $N(E)\propto(E/50{\rm keV})^{\alpha}\times\exp(-E(2+\alpha)/E_{\rm p,o})$ is the cut-off form of the LGRB photon spectrum. The simulation process is described below.

\begin{enumerate}
\item We first simulate 10,000 redshifts ranging from 0 to 10 based on the cumulative distribution function (CDF): \begin{small}
\begin{equation}
\hspace{-10pt}
CDF_{\rm z} = \frac{ \int ^{z_{\rm n}} _{0}  \int^{L_{\rm max}}_{{\rm max}[L_{\rm min},L_{\rm lim}(z)]}\psi(z)\phi(L)\theta\left(P(L,z)\right)dLdz}{\int ^{z_{\rm max}} _{0} \int ^{L_{\rm max}}_{{\rm max}[L_{\rm min},L_{\rm lim}(z)]}\psi(z)\phi(L)\theta\left(P(L,z)\right)dLdz}\;,
\label{eq:CDFz}
\end{equation}
\end{small}
where the intrinsic LGRB redshift distribution $\psi(z)\propto\frac{\psi_\ast(z)(1+z)^\delta}{1+z} \frac{dV(z)}{dz}$ is modeled by the cosmic star formation rate $\psi_\ast(z)$ with a density redshift evolution factor $(1+z)^\delta$, and the intrinsic LGRB luminosity distribution $\phi(L)$ is modeled by a triple power-law function. $\theta\left(P(L,z)\right)$ represents the probabilities of trigger and redshift measurement for current \textit{Swift} LGRBs. $\frac{dV(z)}{dz}$ is the comoving
volume element in a flat $\Lambda$CDM model. The details and fitted parameters of the models can be seen in \cite{2021MNRAS.508...52L}. And the subscript $n$ describes the bins of the CDF.

\item For each mock redshift $z^{\rm mock}_{\rm j}$, where j=1,2, $\cdots$, 10,000, we estimate the corresponding limiting luminosity $L_{\rm lim}(z^{\rm mock}_{\rm j})$, then draw a value from $L_{\rm lim}(z^{\rm mock}_{\rm j})$ to $L_{\rm max}=10^{55}$ erg s$^{-1}$ as the mock luminosity $L^{\rm mock}_{\rm iso, j}$ (or from $L_{\rm min}= 10^{49}$ to $L_{\rm max}=10^{55}$ erg s$^{-1}$, if $L_{\rm lim}(z^{\rm mock}_{\rm j})$ is smaller than $L_{\rm min} = 10^{49}$ erg s$^{-1}$). This behaviour will finally produce 10,000 mock luminosity data. The CDF for luminosity sampling can be written as:
\begin{small}
\begin{equation}
\hspace{-5 pt}
CDF_{\rm L} = \frac{\int^{L_{\rm n}}_{{\rm max}[L_{\rm min},L_{\rm lim}(z)]}\psi(z)\phi(L)\theta\left(P(L,z)\right)dL}{\int ^{L_{\rm max}}_{{\rm max}[L_{\rm min},L_{\rm lim}(z)]}\psi(z)\phi(L)\theta\left(P(L,z)\right)dL}\;.
\label{eq:CDFl}
\end{equation}
\end{small}

\item To obtain the mock observed peak energy $E^{\rm mock}_{\rm p,o}$, we first use $L^{\rm mock}_{\rm iso}$ to get the mock intrinsic peak energy $E^{\rm mock}_{\rm p,i}$ based on the two-dimensional (2D) distribution (\epi, \liso)\footnote{The program used for 2D distributions of data sets can be seen on \url{https://www.mathworks.com/matlabcentral/fileexchange/17204-kernel-density-estimation}} observed by \textit{Swift}, as shown in the upper-right panel in Fig \ref{fig:2D}. Then, taking into account $z^{\rm mock}$, we calculate $E^{\rm mock}_{\rm p,o}$ by $E^{\rm mock}_{\rm p,o}=E^{\rm mock}_{\rm p,i}/(1+z^{\rm mock})$.

\item Similarly, the mock power-law index $\alpha^{\rm mock}$ is obtained through the 2D ($\alpha$, $E_{\rm p,o}$) distribution observed by \textit{Swift}, as shown in the bottom-right panel in Fig \ref{fig:2D}.

\item Finally, we complete the mock LGRB sample with $P^{\rm mock}$ being calculated from $z^{\rm mock}$, $L^{\rm mock}_{\rm iso}$, $E^{\rm mock}_{\rm p,o}$, and $\alpha^{\rm mock}$. This guarantees that each mock burst is an element of a concordant and observable population. The distribution of both mock and observed $P$ are presented in the upper left panel in Fig \ref{fig:1D}.
\end{enumerate}

We need to emphasize that for every parameter, the mock data will only be retained if Kolmogorov-Smirnov (KS) test does not reject the hypothesis that the simulated sample has the same distribution as the observed data in both 1D and 2D\footnote{The program used for KS test for 2D data sets can be seen on \url{https://www.mathworks.com/matlabcentral/fileexchange/38617-kstest_2s_2d-x1-x2-alpha}} \citep{1983MNRAS.202..615P} data sets with a 5\% significance level. For the \textit{Swift} data (up to GRB 240523A, there are 372 LGRBs having measured redshift in the \textit{Swift} sample), to reduce the uncertainty caused by the estimation of spectral parameters, only LGRBs with $15 \leq E_{\rm p,o} \leq 9000$ (keV) are included, leaving 292 LGRBs. Similarly, we only use LGRBs with $P\ge0.5$ ph cm$^{-2}$ s$^{-1}$ to avoid the uncertainty at the very faint end where the data are highly incomplete (see similar operation in \cite{2021A&A...649A.166P}, and the low probabilities of \textit{Swift} LGRB detection trigger and redshift measurement at faint $P$ in \cite{2021MNRAS.508...52L}). At the end, there are 259 bursts included in the observed sample used for this work. Note that the limiting luminosity in Eq. \ref{eq:CDFz} and \ref{eq:CDFl} is determined by the peak-flux limit $P_{\rm lim}=0.5$ ph cm$^{-2}$ s$^{-1}$ of the observed data.

\section{Result and discussion}
\label{sec:result}

\begin{figure*}
\center
\includegraphics[angle=0,scale=0.6]{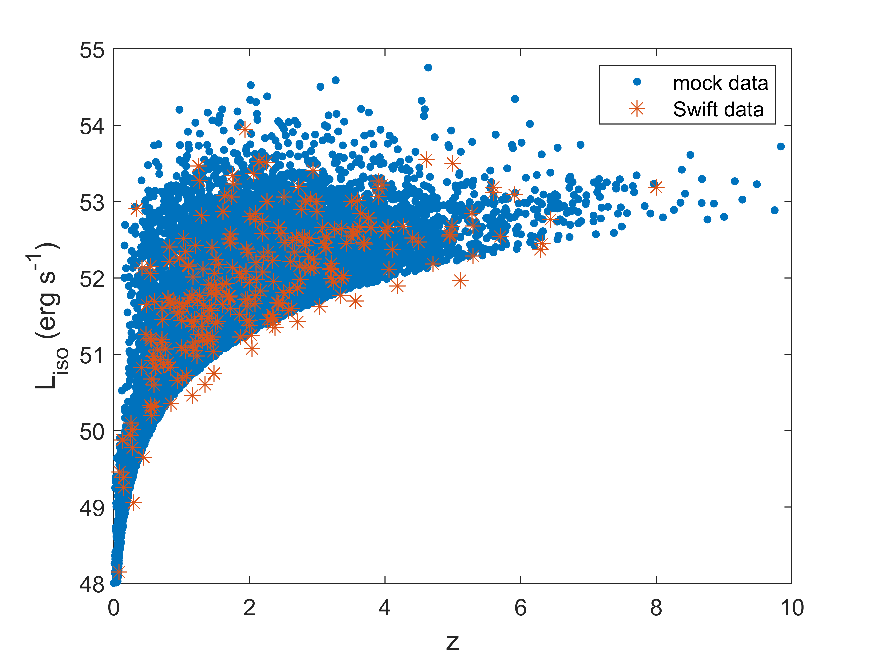}
\includegraphics[angle=0,scale=0.6]{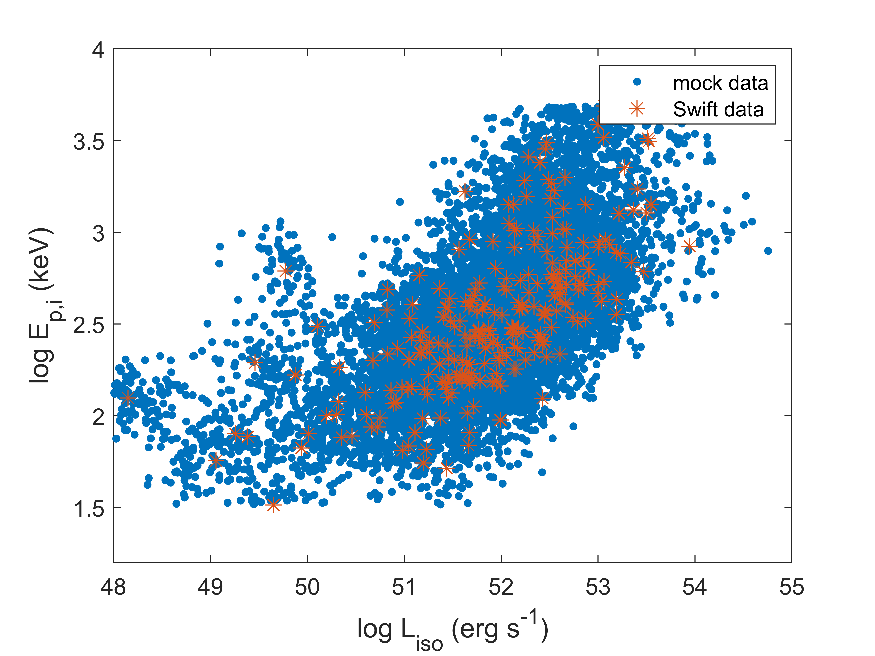}
\includegraphics[angle=0,scale=0.6]{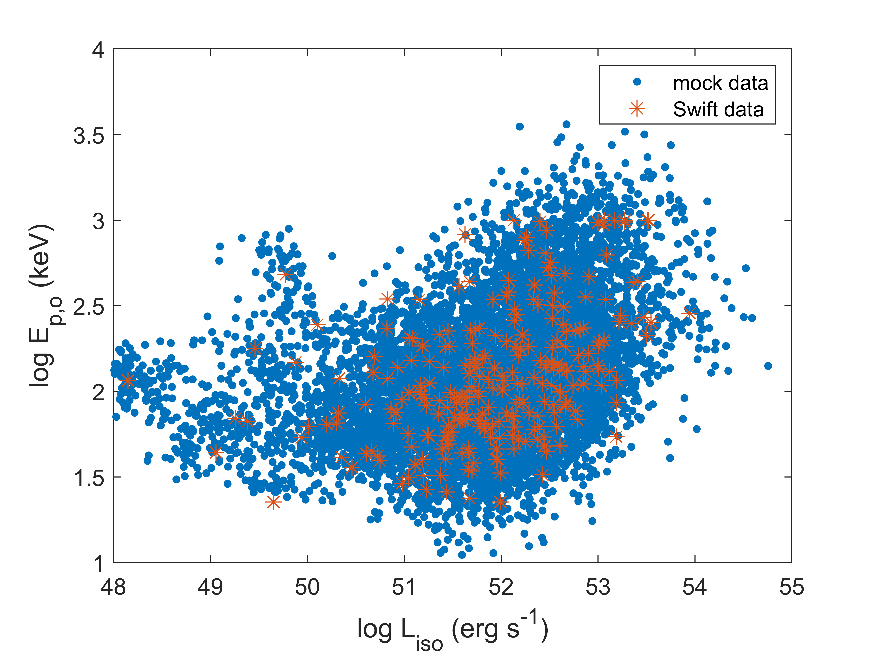}
\includegraphics[angle=0,scale=0.6]{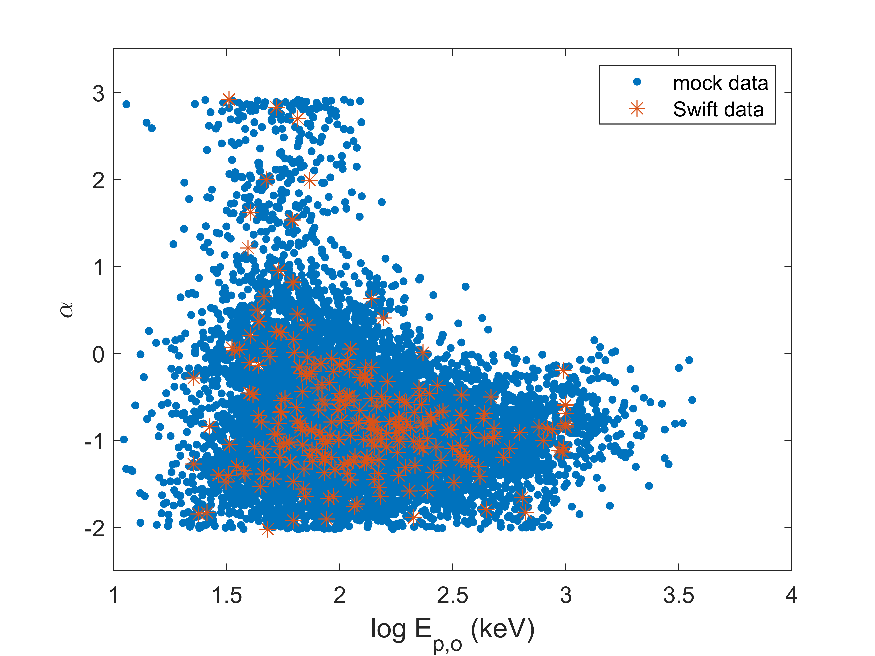}
\caption{The joint distributions of parameters \liso\ vs. $z$ (upper-left), \epi\ vs. \liso\ (upper-right), $E_{\rm p,o}$ vs. \liso\ (bottom-left) and $\alpha$ vs. $E_{\rm p,o}$ (bottom-right) for both simulated data (blue points) and observed data (red star symbols). In the upper-left panel, some observed points are located below the lower limit of simulated luminosity because the limiting luminosity is estimated with average spectral parameters which can only reflect an average luminosity limit. However, the null hypothesis that both simulated and observed data sets were drawn from the same continuous distribution is not rejected by the KS test with a 5\% significance level in each diagram.
\label{fig:2D}}
\end{figure*}

\begin{figure*}
\center
\includegraphics[angle=0,scale=0.6]{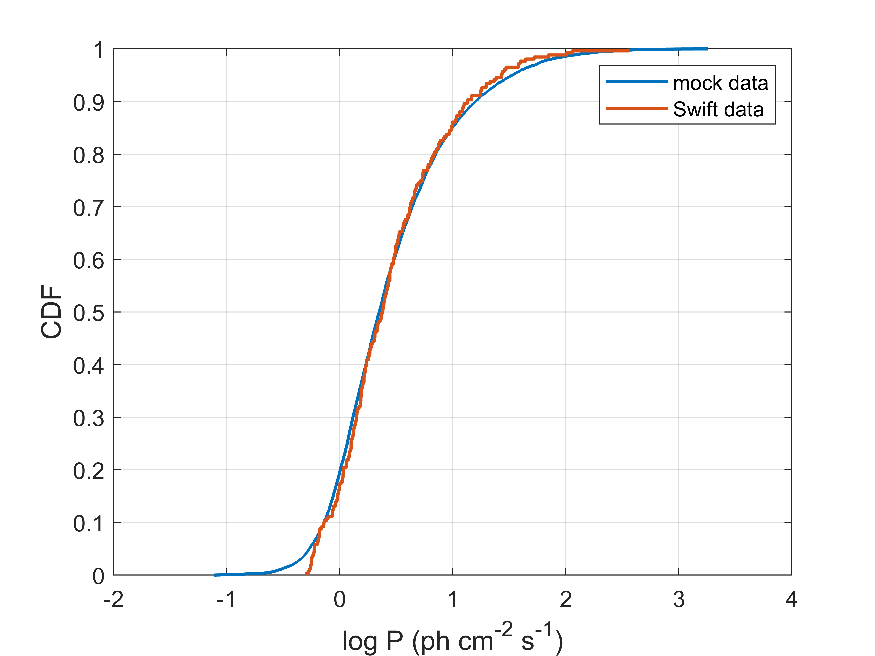}
\includegraphics[angle=0,scale=0.6]{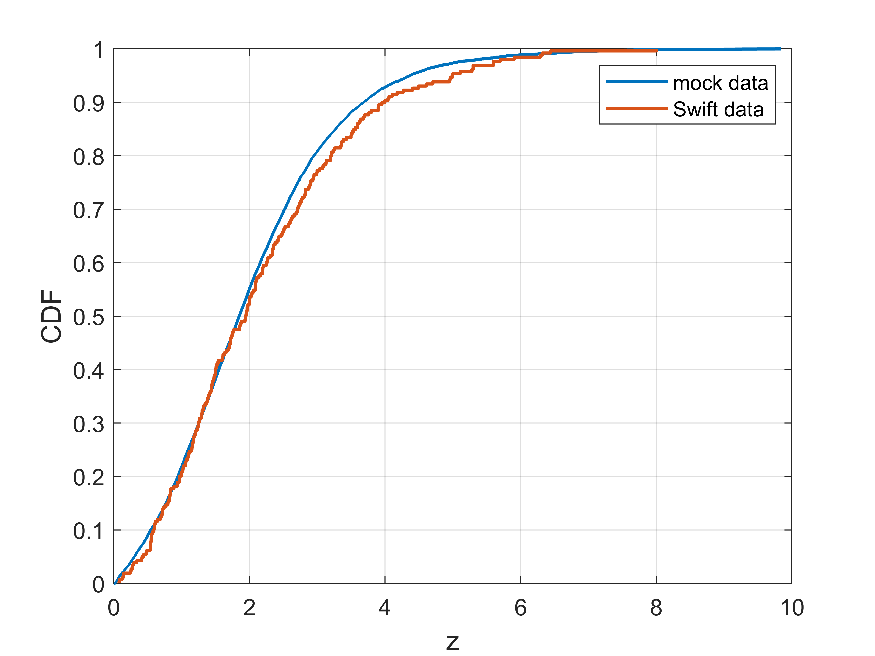}
\includegraphics[angle=0,scale=0.6]{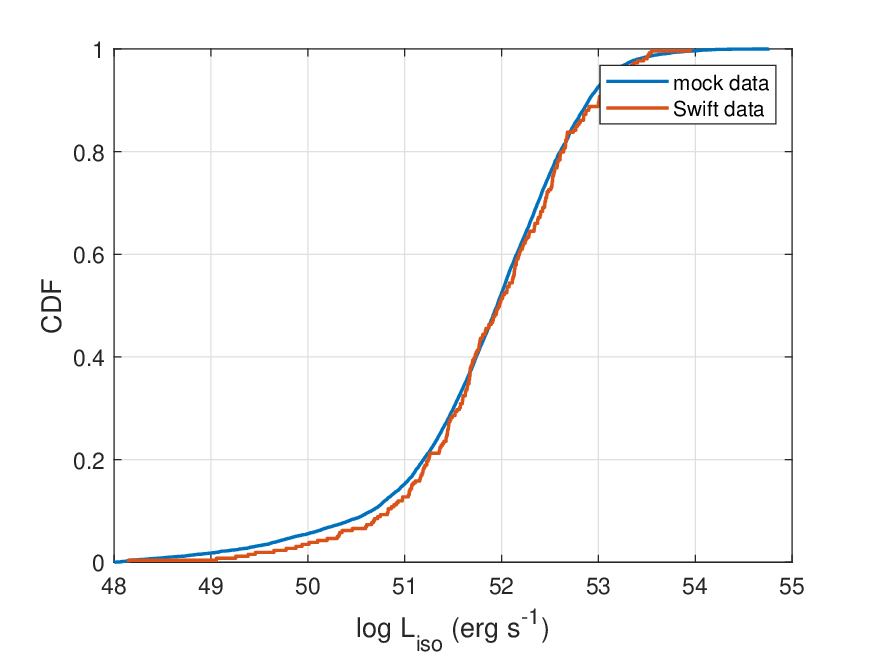}
\includegraphics[angle=0,scale=0.6]{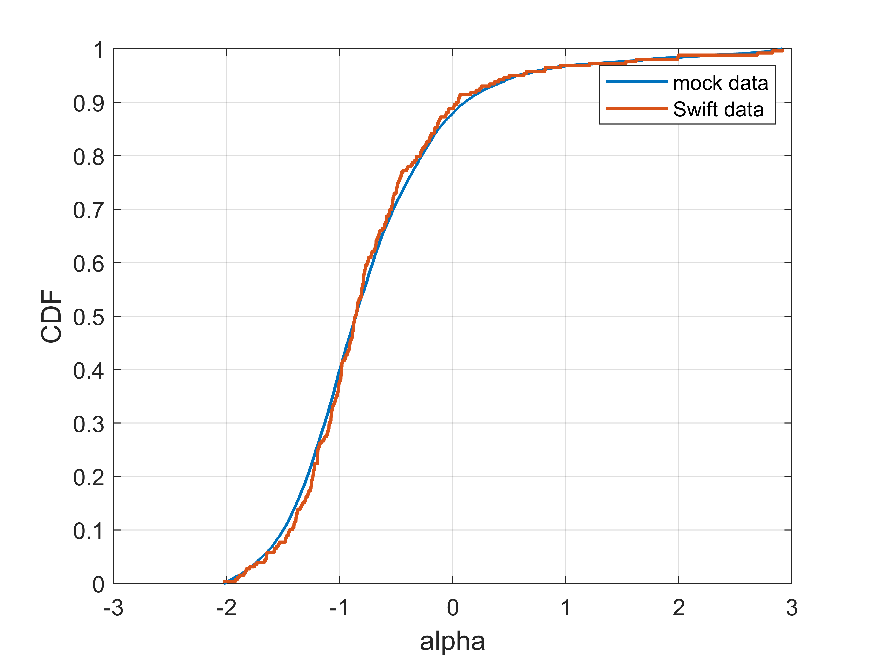}
\includegraphics[angle=0,scale=0.6]{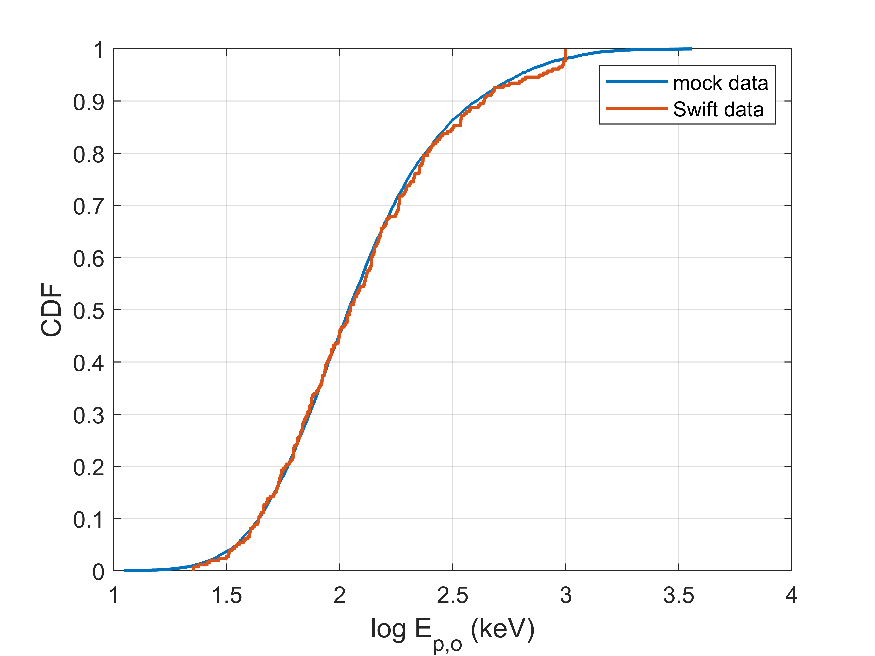}
\includegraphics[angle=0,scale=0.6]{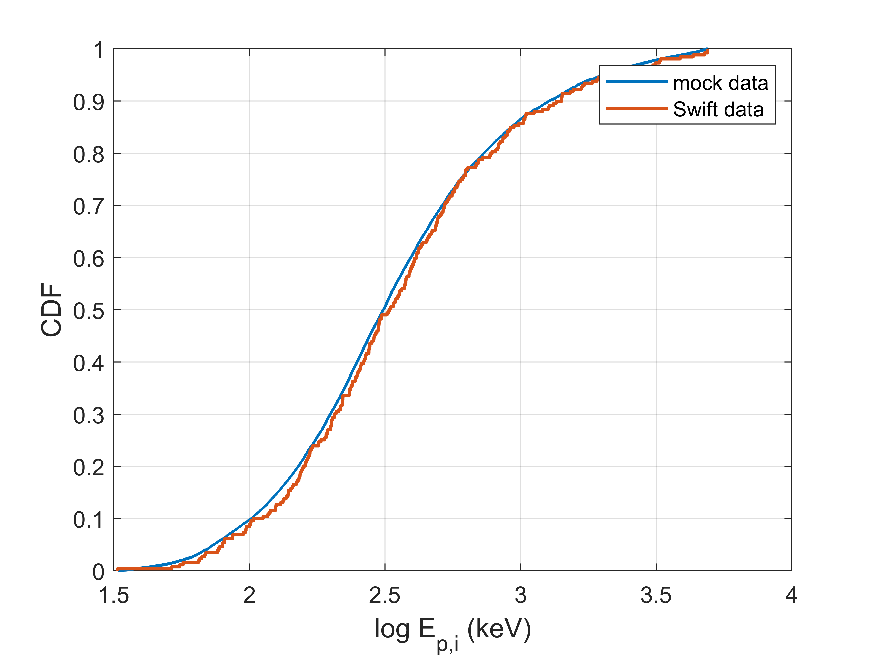}
\caption{The cumulative distributions of peak flux $P$, redshift $z$, luminosity \liso, and spectral power-law index $\alpha$, observed peak energy $E_{\rm p,o}$, and rest-frame peak energy \epi\ of LGRBs. The blue curves represent the mock data, and the red curves represent the observed data from \textit{Swift} catalog. In each diagram, the null
hypothesis that both simulated and observed data sets were drawn from the same continuous distribution is not rejected by the
KS test with a 5\% significance level
\label{fig:1D}}
\end{figure*}

The two-parameter diagrams for both simulated and observed data are presented in Fig \ref{fig:2D}. It is clear that the data points of the simulated samples have distributions well consistent with the \textit{Swift} data points, which is also supported by the KS test used for 2D data sets. In Fig \ref{fig:1D}, the mock data are also verifiably in agreement with the corresponding observed distributions for each parameter. These illustrate that our simulated data provide a realistic representation of the \textit{Swift} observed population.

With the simulated sample, we study the reality of the relationship between \epi\ and \liso , and the impact of $P$-selection on the \epi--\liso\ correlation. We finally propose a tool based on the \epi--\liso\ diagram, to quickly identify potential LGRBs that may not come from collapsars. The details are described in the following.

\subsection{On the reality of the \epi--\liso\ correlation}
\label{ssec:result1}

The mock sample used here takes into account the correlations between pairs of LGRB parameters. However, during the simulation process, we realized that, even though it is the observed peak energy \epo\ that is required to simulate the sample, the observed $P$ distribution can not be reproduced well if we only consider the possible dependence of \epo\ on other parameters. To get mock $P$ data that follow the observed distribution, it is mandatory to include in the simulation the correlation between the intrinsic peak energy \epi\ and the LGRB luminosity \liso . This indicates that the correlation of these two parameters is a key intrinsic property of LGRBs, a finding which was also pointed out by \cite{2021A&A...649A.166P}, among others. A physical link between \epi\ and \liso\ may be invoked to explain this dependence, which, on another hand, supports a physical interpretation of the boundaries of the regions occupied the different types of transients (LGRBs, SGRBs, SGR Giant Flares) in the (\epi, \liso) diagram.

\subsection{Impact of the peak flux}
\label{ssec:result2}
\begin{figure}
\center
\includegraphics[angle=0,scale=0.65]{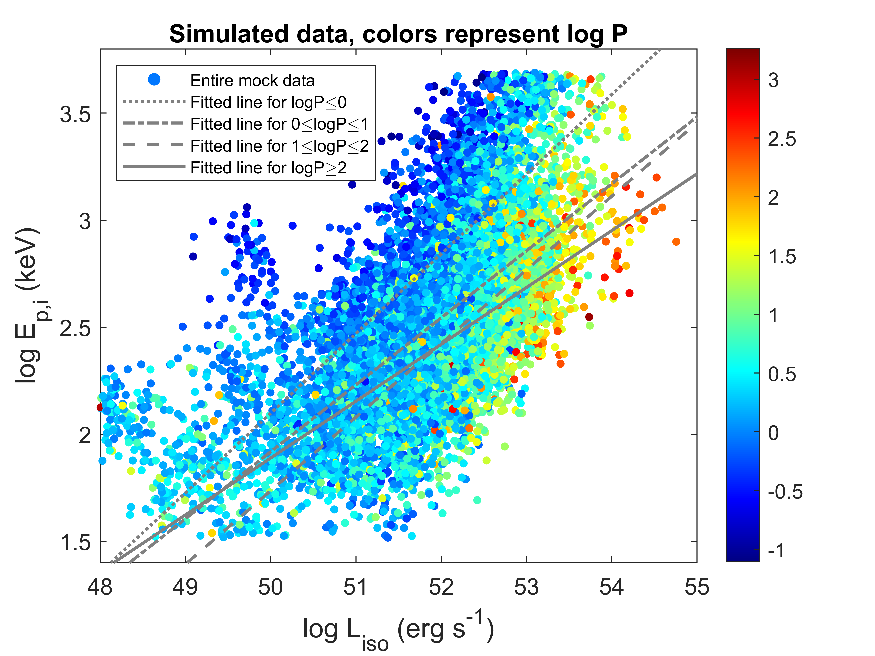}
\caption{The \epi\ vs. \liso\ distribution for the mock data. The colors represent the different levels of $\log P$ values. The four gray lines are best-fit \epi--\liso\ relations for the entire mock data with $\log P\le0$ (dotted), $0\le\log P\le1$ (dash-dotted), $1\le\log P\le2$ (dashed) and $\log P\ge2$ (solid), respectively.}
\label{fig:epilp}
\end{figure}

\begin{figure}
\center
\includegraphics[angle=0,scale=0.65]{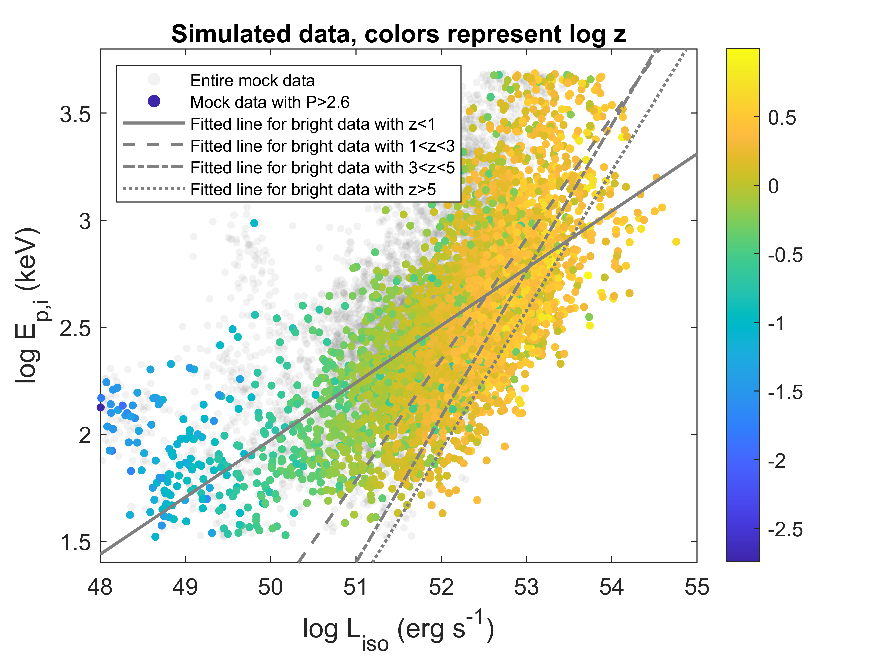}
\caption{The \epi\ vs. \liso\ distribution for the mock data. Data with $P>2.6$ are highlighted with colors representing the different levels of $\log z$ values. The four gray lines are best-fit \epi--\liso\ relations for bright data with $z<1$ (solid), $1<z<3$ (dashed), $3<z<5$ (dash-dotted), and $z>5$ (dotted), respectively. The best-fit parameters ($a$, $b$) with 1$\sigma$ uncertainty in each $z$ group are ($0.27\pm0.01$, $-11.39\pm0.34$), ($0.57\pm0.01$, $-27.24\pm0.46$), ($0.68\pm0.01$, $-32.97\pm0.81$), ($0.65\pm0.06$, $-31.97\pm3.62$), respectively.
\label{fig:epilz}}
\end{figure}

To reveal the impact of $P$ on the \epi--\liso\ correlation, we show in Fig \ref{fig:epilp} the \epi\ vs. \liso\ data with colors representing the $\log P$ values. Here we can see that the boundaries for data points with different colors (i.e. different values of $\log P$) are significantly different. For instance, the regions of blue points (low $\log P$) and red points (high $\log P$) have only a faint overlap. We estimate the best-fit \epi--\liso\ relations for data within different $\log P$ ranges and show the best-fit parameters in Table \ref{tab1} and the best-fit lines in Fig \ref{fig:epilp}. A discrepancy is clear between the lines coming from different peak flux bins.

\begin{table*}
\centering
\caption{Best-fit parameters of \epi --\liso\ relations in simulated data with different $P$ ranges}
\label{tab1}
\begin{tabular}{ccccc}
\hline
  $P$ ranges & $a$ (entire data) & $b$ (entire data) & $a$ (high-\liso\ data) & $b$ (high-\liso\ data) \\
\hline
  $\log P\le0$ & $0.37\pm0.01$ & $-16.52\pm0.28$& $0.68\pm0.01$ & $-32.59\pm0.29$\\
  $0\le\log P\le1$ & $0.31\pm0.01$ & $-13.73\pm0.10$& $0.49\pm0.01$ & $-22.81\pm0.13$\\
  $1\le\log P\le2$ & $0.34\pm0.01$ & $-15.46\pm0.22$& $0.44\pm0.01$ & $-20.49\pm0.25$\\
  $\log P\ge2$ & $0.27\pm0.02$ & $-11.39\pm0.69$& $0.33\pm0.02$ & $-15.03\pm0.77$\\
  \hline
\end{tabular}
\begin{description}
  \item[\emph{Note.}] {The relation formula is considered as $\log \frac{E_{\rm p, i}}{1\ \rm keV}=a\log \frac{L_{\rm iso}}{1\ \rm erg/ s}+b$. The first columns of $a$ and $b$ results come from entire mock data, and the second columns of $a$ and $b$ results come from mock data with $L_{\rm liso}\ge10^{50.5}$ erg s$^{-1}$. Errors show the 1$\sigma$ uncertainty of the corresponding parameters.}
\end{description}
\label{tab1}
\end{table*}

Fig \ref{fig:epilz} provides another view on the impact of $P$. Since $P>2.6$ (ph cm$^{-2}$ s$^{-1})$ data have $\sim44\%$ fraction of the entire mock sample, we highlight the data points with $P>2.6$ ph cm$^{-2}$ s$^{-1}$ in the \epi\ vs. \liso\ panel by giving colors to these points reflecting the values of $\log z$. It shows that bright bursts are concentrated in the lower part of the full simulated population. It is another illustration of the fact that samples constrained by different $P$ ranges lead to different \epi --\liso\ correlations.

Additionally, since every boundary is not dominated by one single value of redshift, the edges of (\epi, \liso) data should not be determined by the limit of present redshift-measurement tools. We can also notice that data points with high-$z$ generally have large values of \epi\ and \liso . This is an observational effect. First, the detection of high-$z$  - low-\liso\ GRBs is limited by the instrumental sensitivity, leading to the lack of such events in current samples. Second, the lack of GRBs with high-\liso\ at low-$z$ is explained by the rarity of such GRBs, which can only be detected in large volumes of the universe \citep{2010MNRAS.406.1944W,2012ApJ...749...68S,2016A&A...587A..40P,2021MNRAS.508...52L}. These observational effects may be the main cause to the finding that LGRBs in different redshift groups have different best-fit \epi --\liso\ relation in previous works \citep[e.g.][]{2007MNRAS.379L..55L} as well as in this work. In Fig \ref{fig:epilz}, we show the best-fit lines for bright mock data in different $z$ bins. For data with $z>1$, the lines in different $z$ groups are discrepant but nearly parallel. The shift of lines and the evolution of limiting luminosity with redshift are consistent. It is believed that the evolving \epi --\liso\ relations are probably caused by the varying limiting luminosity in different $z$ bins.

\subsection{GRBs in the low-\epi\ \& \liso\ region}
\label{ssec:result3}
The best-fit line for $z<1$ data is extremely different from that in the other $z$ groups. This is associated with the strange $P$ distribution in the low-\epi, low-\liso\ region. The colors of the points in Fig \ref{fig:epilp} suggest that the $P$ distribution in the low-\epi, low-\liso\ region (the lower left corner of the plot) is not the simple extrapolation of the $P$ distribution in the high-\epi, high-\liso\ region. Specifically, the low-\epi, low-\liso\ region contains many GRBs with high peak flux (green, orange, and red points), which prefer to occupy the region with higher values of \epi\ and \liso.

As also shown in Table \ref{tab1}, getting the \epi--\liso\ relations from the mock data with $L_{\rm liso}\ge10^{50.5}$ erg s$^{-1}$, we find that the slopes evolve simply with $P$, unlike the highly variable slopes measured for the entire mock data. This also accounts for the strange $P$ distribution in the low-luminosity region.

Therefore, we hypothesize the existence of a subgroup of LGRBs in this region, leading to this discrepancy. To simply address this possibility, we select observed bursts with $L_{\rm iso}\le10^{50}$ erg s$^{-1}$, $E_{\rm p, i}\le10^{2.5}$ keV, and $P\ge10^{0.5}$ ph cm$^{-2}$ s$^{-1}$ (lower value for green points in Fig. \ref{fig:epilp}), and find that GRB 060614, 161219B, 191019A, and 230328B survive this selection. Of these bursts, the origins of GRB 060614 \citep{2006Natur.444.1044G} and GRB 191019A \citep{2023NatAs...7..976L} are believed to be mergers, which suggests the \epi--\liso\ diagram with $P$ could be used as an indicator to point to LGRBs with non-collapsar progenitors.

\subsection{The typical \epi\--\liso\ relation \& The excess of simulated data}
\label{ssec:result4}
In the above, we study the \epi\--\liso\ relation in different $P$ and $z$ bins to investigate the selection effects from these properties. As discussed above, the peak-flux selection, the limiting luminosity, and the potential sub-class LGRBs in the low-\epi, low-\liso\ region may have a significant impact on the best-fit results of the \epi\--\liso\ relation, which leads most of our best-fit \epi\--\liso\ relations to be different from previous findings. However, we notice that our results with data at middle \liso\ and $z$ ranges are similar to the typical \epi\--\liso\ relation. This implies that the previous findings are probably dominated by the subset of luminous LGRB with moderate or large peak flux, preventing obtaining a comprehensive perspective on the entire LGRB population.

Although our simulated sample has proven to be well compatible with the observed sample, some remaining differences deserve further discussion. For instance, the upper limit of mock \epo\ is greater than the Swift \epo\ data which is constrained by 1000 keV for the automatic fit for \textit{Swift} LGRB spectra, and mock $P$ with values smaller than $P_{\rm lim}$ also occupies 4\% of the entire mock data. For the \textit{Swift} high-\epo\ data, we notice that there is an unnatural jump at the bright end caused by the preset limit for the spectral fitting. However, since the fraction of this part of \epo\ data is merely $\sim2$\% and the mock $\alpha$ is well consistent with the observed results, the conclusions of this work are hardly affected. We can confirm, based on the 2D (\epo, $P$) and ($P$, $z$) distributions, that the excess of the faint mock $P$ is scarcely associated with the high-\epo\ data, but mainly associated with the middle values of mock $z$. After removing mock data with the faint $P$, the mock $z$ and (\epi, \liso) distributions are still in good agreement with the observed results, and our conclusions remain valid.

\section{Summary}
\label{sec:sum}

In this work, we study the potential selection effects in the LGRB \epi--\liso\ correlation, based on a mock population representative of the \textit{Swift} GRB sample. We obtain the mock $z$ and \liso\ data from the LGRB rate and luminosity functions in \cite{2021MNRAS.508...52L}, and the mock $E_{\rm p,o}$ and $\alpha$ from the observed (\epi, \liso) distribution and ($\alpha$, $E_{\rm p,o}$) distribution respectively. Finally, we calculate $P$ through these mock $z$, \liso, $E_{\rm p,o}$, and $\alpha$ data. All simulated single-parameter distributions, and two-parameters joint-distributions are well consistent with the observational results.

The simulated data show that the (\epi, \liso) distribution is partly shaped by the value of $P$. This impact is also reflected in the best-fit (\epi, \liso) relation, as demonstrated by the best-fit parameters measured for mock GRB samples with different peak flux. This means that a detailed analysis of the impact of the peak flux selection is necessary when using the \epi-- \liso relation.

Finally, the $P$ distribution at low-\epi\ \& \liso\ region appears different from the simple extrapolation of the distribution in the higher-\epi\ \& \liso\ region, a fact which also affects the slopes of the best-fit relations. One simple interpretation is the existence of a subgroup of LGRBs in the low-\epi\ \& \liso\ region that affects the $P$ distribution, as  suggested in previous works \citep[see e.g.][]{2015ApJ...806...44P}, which may also link to the triple-power-law LGRB luminosity function reported in \cite{2021MNRAS.508...52L}. To get a simple proof of this idea, we pick out bursts with $L_{\rm iso}\le10^{50}$ erg s$^{-1}$, $E_{\rm p, i}\le10^{2.5}$ keV, and $P\ge10^{0.5}$ ph cm$^{-2}$ s$^{-1}$ and find that GRB 060614 and 191019A, which have low probabilities of being associated with SNe, are included in the sample selected. This suggests that the \epi--\liso\ diagram with $P$ taken into account may be used to point to LGRBs from other origins.

Note that our simulated sample needs to be take into account an instrinsic (\epi, \liso) correlation; otherwise, the simulated peak flux distribution cannot be made consistent with the observed distribution. This illustrates the physical basis of the (\epi, \liso) distribution. In particular, the boundary on the bright GRB side, which limits the occurrence of GRBs with high-\liso\ \& low-\epi . To study the reliability of this limit, we specially select data with $P>2.6$ (ph cm$^{-2}$ s$^{-1}$) from the entire mock sample. We find that this limit is not dominated by the minimum or maximum LGRB redshift and peak flux, which also means the limit is not due to observational effects.

In view of the broad usage of the \epi--\liso\ relation to constrain the LGRB emission mechanisms, to assist LGRB as probes in the early universe, and to better understand the GRB origins, it is important to get the most complete view of the distribution of GRBs in the \epi--\liso\  plane.
In this respect, the detection of GRBs with lower $E_{\rm p}$ than the current population will be crucial to clarify the position of low-\epi\ and low-luminosity LGRBs in the \epi--\liso\ diagram of LGRBs. Fortunately, the Einstein Probe \citep{2022hxga.book...86Y} and SVOM \citep{2016arXiv161006892W,2022IJMPD..3130008A} satellites, which have been recently launched, have shown to expand the GRB detection to $\sim$ few keV while benefiting from excellent followup with large telescopes and a high successful rate for redshift measurement. A more holistic view of the (\epi, \liso) data and the GRB classification may thus be supported by SVOM and Einstein Probe observational results.

\begin{acknowledgments}
We thank all colleagues who contributed to the \textit{Swift} GRB catalog.
\end{acknowledgments}

\bibliography{sample631}{}
\bibliographystyle{aasjournal}



\end{document}